\documentstyle[12pt,aasms4]{article}

\begin{document}

\title{A ROSAT Deep Survey of Four Small Gamma-Ray Burst Error Boxes}

\author{K. Hurley, P. Li} 
\affil{University of California at Berkeley, Space Sciences Laboratory, CA 
94720-7450}
\authoremail{pli@sunspot.ssl.berkeley.edu}

\author{M. Boer}
\affil{CESR, B.P. 4346, 31029 Toulouse Cedex, France} 

\author{T. Cline}
\affil{Goddard Space Flight Center, Code 661, Greenbelt, MD, 20771}

\author{G. J. Fishman, C. Meegan}
\affil{Marshall Space Flight Center ES 62, Huntsville, AL 35812}

\author{C. Kouveliotou}
\affil{Universities Space Research Association, Marshall Space
Flight Center ES-84, Huntsville, AL 35812}

\author{J. Greiner}
\affil{MPE, D-85740 Garching, Germany}

\author{J. Laros}
\affil{University of Arizona, Department of Planetary Sciences, Tucson, AZ  85721}

\author{C. Luginbuhl, F. Vrba}
\affil{US Naval Observatory, Flagstaff Station, P.O. Box 1149, Flagstaff, AZ 86002-1149}

\author{T. Murakami}
\affil{ISAS, 3-1, Yoshinodai, Sagamihara, Kanagawa 229, Japan}

\author{H. Pedersen}
\affil{Copenhagen University Observatory, Oster Voldgade 3, DK 1350, 
Copenhagen, Denmark}

\author{J. van Paradijs\altaffilmark{1}}
\affil{University of Alabama in Huntsville, Huntsville, AL 35899}

\altaffiltext{1}{Astronomical Institute 'Anton Pannekoek', University of
Amsterdam, Kruislaan 403, 1098SJ Amsterdam, The Netherlands}

\begin{abstract}

We have used the ROSAT High Resolution Imager (HRI) to search for quiescent X-ray counterparts to four
gamma-ray bursts (GRBs) which were localized to small ($\rm \leq 10\, arcmin.^2$)  
error boxes with the Interplanetary Network (IPN).  The observations
took place years after the bursts, and the effective exposure times for 
each target varied from $\sim$ 16 - 23 ks.  We have not found any X-ray
sources inside any of the error boxes. The 0.1 - 2.4 keV 3$\sigma$ flux upper limits range
from around $\rm 5\times10^{-14}\, erg\, cm^{-2}s^{-1}$  to $\rm 6\times10^{-13} erg \,cm^{-2}s^{-1}$ depending on the burst and the assumed shape of the quiescent spectrum. 
We consider four types of X-ray emitting galaxies (normal, AGN, faint, and star-forming) and
use the flux upper limits to constrain their redshifts.  We then use the
GRB fluences to constrain the total energies of the bursts.

\end{abstract}


\section{Introduction}

The cosmic gamma-ray burst distance scale was a mystery until the recent 
discovery of  X-ray afterglows by BeppoSAX, and of optical and radio afterglows from some of these GRBs. These observations have made it possible 
to estimate the GRB distance scale from the redshift of the optical counterparts. Four 
spectroscopic redshift 
measurements have now been obtained: z=0.835 for  GRB970508 (Metzger et al. 1997;
Bloom et al. 1998a),  z = 
3.4 for GRB971214 (Kulkarni et al. 1998), z= 0.966 for GRB980703 (Djorgovski
et al. 1998), and z=1.6 for GRB990123 (Hjorth et al. 1999).  The nature of the host galaxies in these cases is not clear, although
in one case, the host appears to be in an active star-formation phase (Djorgovski et al.
1998).  While it is now generally accepted 
that GRBs are at cosmological distances, the picture is clouded by the apparent
association of one burst, GRB980425, with a nearby (z=0.008) supernova, SN1998bw, in
a barred spiral galaxy, ESO184-G82 (Galama et al. 1998). 
Although there is evidence that
GRBs are not \it generally \rm associated with supernovae (Kippen et al. 1998),
it has been argued that perhaps 1\% could be (Bloom et al. 1998b).  Similarly, it
has been argued that GRBs are associated with Abell clusters (Kolatt and Piran 1996;
Struble and Rood 1997) and radio-quiet quasars (Schartel et al. 1997).  While
the latter two associations are probably not valid (Hurley et al. 1999), all these claimed
associations serve to demonstrate that the nature of GRB host galaxies is not yet well understood.
Indeed, the evidence is not inconsistent with the suggestion that short gamma-ray bursts have a different origin from
long ones (Pizzichini 1995; Belli, 1997; Tavani 1998; but see also Kouveliotou et al. 1996
and Pendleton et al. 1997).   Counterpart observations in the x-ray, optical, and radio ranges at various times after 
the bursts will continue to be valuable for unraveling these issues.

In this paper, we present ROSAT soft X-ray observations of four GRB error boxes.  The objective
of this study was to detect quiescent X-ray counterparts to GRB sources.  Accordingly,
the observations
took place years after the bursts.  (At the sensitivity levels of this survey,
the fading X-ray counterparts to these bursts would have been undetectable
days to a week after the bursts.)  The GRB locations were obtained with the Interplanetary 
Network (IPN), composed of \it Ulysses \rm, BATSE aboard the \it Compton Gamma
Ray Observatory \rm, and either the \it Pioneer Venus Orbiter \rm 
or \it Mars Observer \rm spacecraft at the times of these 
bursts.  The error boxes have been published in Laros et al. (1997, 1998), and
the properties of the bursts have appeared in Meegan et al. (1996).  
The bursts were selected for study based on three criteria: their high galactic
latitude and/or the error box area (from which the probability
of detecting a random source within the error box can be estimated),
and the maximum error box dimension (so that the error box could be covered
in a single ROSAT High Resolution Imager pointing).  

\section{ROSAT Results} 

Table 1 gives the properties of the bursts and the results of the observations.
The BATSE trigger number (Meegan et al. 1996) appears in the first row, and the
ROSAT sequence number in the second.
The galactic latitude
and error box area are given in the next two rows.  
Row 5 gives the approximate 25 - 150 keV fluence, estimated from
the \it Ulysses \rm observations; all of the bursts are rather bright (or they
would not have been detected by the relatively small instruments of the IPN), and thus,
presumably, relatively nearby.  The following row gives the total effective
observation time by ROSAT.  All observations took place with the High Resolution
Imager, in the 0.07 - 2.4 keV energy band.  Row 7 gives the elapsed time between
the burst and the observation.  In this particular study, no attempt was
made to minimize this time interval, as the objective was the study of long-lived, quiescent
counterparts.  Row 8 indicates whether a source was observed within the
IPN error box.  Row 9 gives the number of sources in the HRI field of view.
Row 10 gives the \it a posteriori \rm probability of finding a source in
the error box by chance.  It is calculated simply as the number of sources
times ratio of the error box area to the area of the HRI field of view.    
(The number of sources is consistent with the number of background sources
expected at these sensitivity levels.)  Row 11
gives the hydrogen column density along the line of sight.  The last
3 rows give the upper limits to the 0.1 - 2.4 keV flux of any source in the error box
assuming thermal bremsstrahlung, blackbody, and power law spectra, corrected
for the foreground column density.  Depending on the column density and the
assumed spectral form, different spectral parameters may result in upper limits
which differ by a factor two to three.

The ROSAT data were analyzed using the standard data analysis tools and
data cuts.  In these observations, there is presently a boresight error that is generally less than 10 $\arcsec$.  As this is much smaller than the  error box sizes, and since the boresight-corrected data
were not all available at the time of this writing, we have ignored this.  The data selection criteria are given in Gruber et al. (1996). The 
IRAF/PROS sliding window source detection technique and software were used
to identify the 
sources in each image (we used a signal to noise ratio of 3 for our source
detection criterion).  To obtain the 3 $\sigma$ upper limits, the source counts and background  counts were taken from the same image. The source box center was at the GRB error box center.  The background box was centered at a location where it excluded the GRB error box and any
sources in the field. The box size was initially chosen to be $\rm32\arcsec \times 32\arcsec$.  The ROSAT HRI point spread function is such that this box contains 90$\%$ of  the energy of a point source.  The box size was the same for both source and background estimates.  The column density for each GRB error box was obtained 
from the HEASARC. \footnote{http://heasarc.gsfc.nasa.gov/docs/frames/hhp\_sw.html} The HRI images with the GRB error boxes are shown in figures 1-4.

The 3 $\sigma$ count upper limit was then converted to flux for three assumed spectral models since the HRI does not give spectral information.  For the thermal bremsstrahlung and blackbody models, a temperature kT = 1 keV was assumed. For the power law model, the assumed photon spectral index was -1. With the kT or spectral index and the line-of-sight column density, we
used PIMMS (Portable Interactive Multi-Mission Simulator, available through the HEASARC
\footnote{http://heasarc.gsfc.nasa.gov/Tools/w3pimms.html}) to convert the 3 $\sigma$ count rate upper limit to a 0.1-2.4 keV 3 $\sigma$ flux upper limit. 

\section{Constraints on GRB host galaxies}
	
All of the bursts in Table 1 would probably be classified as "long".  One possible
exception is GRB920325.  No duration is given for this event in the BATSE
catalog (Meegan et al. 1996).  The burst consists of a short ($<$ 1 s) initial
spike, but is followed by low-level emission for several seconds.  The next
shortest event, GRB930706, has a T$_{90}$ duration of 2.7 s, placing it between
the "short" and "long" peaks of the duration distribution, but on the shoulder
of the "long" bursts.  We therefore assume that these GRBs are similar to the
four for which redshifts have been measured, and that they originated in or
very close to galaxies.
The X-ray flux upper 
limits may therefore be used to derive lower limits to the distances of
the hosts.  From this, in turn, we may obtain lower limits 
on the GRB energy.  

\subsection{Galaxy types and distance scale} 

Galaxies display a wide range of X-ray luminosities.  The X-rays from normal
galaxies come from individual sources such as binaries and supernova remnants, as
well as from a hot phase of the interstellar medium, heated by supernovae;
their X-ray luminosities range from $\rm L_x = 10^{38}\, to 10^{42}\, erg\, s^{-1}$ 
(Fabbiano 1989). The X-ray emission from AGN's is thought to be powered by supermassive
black holes; AGN X-ray luminosities range from $\rm L_x = 10^{42}\, to 10^{46}\, erg\, s^{-1}$
(Maccacaro et al. 1991).  Perhaps more relevant to the host galaxies of gamma-ray bursts,
studies of faint galaxies (B $\leq$ 23) at redshifts 0.1 $<$ z $<$ 0.5 show
that they have
X-ray luminosities $\rm L_x = 10^{41.5}\, to 10^{43}\, erg\, s^{-1}$ (Roche et al. 1995).
These rather large X-ray luminosities are apparently not due to a Malmquist bias,
since their $\rm L_x/L_B$ ratios are an order of magnitude larger than most local
galaxies.
Finally, studies of star-forming galaxies, whose X-ray emission may be due to massive X-ray
binaries, indicate that their luminosities are $\rm L_x = 10^{39.5}\, to 10^{41.5}\, erg\, s^{-1}$
(Griffiths and Padovani 1990).

In an $\rm \Omega=1, \Lambda=0$ universe, the quiescent X-ray luminosity L$_0$, 
the luminosity distance d$_L$, and the observed X-ray flux F$_x$ are related through 
\begin{equation}
F_x=\frac{L_0 (1+z)^{-\alpha}}{4\pi d_L^2}
\end{equation}
where z is the redshift, and the photon spectrum of the quiescent X-ray source is assumed to be a power law with
photon index $\alpha$.  The luminosity distance is given by
\begin{equation}
d_L=\frac{2c(1+z-\sqrt(1+z))}{H_0}
\end{equation}
where c is the speed of light and H$_0$ is the Hubble constant. Equations 1 and 2
may be used to derive a lower limit to the redshift of each of the four bursts, assuming
a specific galaxy type.  From this, the total gamma-ray energy in the burst may be calculated:
\begin{equation}
E_\gamma = 4\pi f d_L^2 (1+z)^n
\end{equation}
where f is the burst fluence, and the spectral shape of the burst is assumed to
be a power law with photon index n.  In table 2, distance lower limits to each
burst are given for the four galaxy types, for the lower and upper limit to the
X-ray luminosity.  A lower limit to the total isotropic burst energy is also calculated.  We have assumed
that the X-ray spectrum is given by a power law with $\rm \alpha=-1$ and that
the GRB spectrum is described by a power law with $\rm n=-2$, and have taken
H$_0$=65 km s$^{-1}$ Mpc$^{-1}$.  Again, different assumptions about spectral
parameters, as well as cosmological constants, will result in different limits.

\section{Discussion}

Numerous multi-wavelength follow-up observations have been performed on the four
GRB error boxes in this study, including Barthelmy et al. (1994) (Schmidt telescope observations); Luginbuhl et al. (1995) (UBVI observations at
the USNO); Hurley et al. (1995) (extreme ultraviolet observation
with the \it EUVE \rm spacecraft); Luginbuhl et al. (1996) (optical observations
at USNO and CTIO); Schaefer et al. (1997) 
(\it Hubble Space Telescope \rm observations in the B, U, and UV bands);
Larson and McLean (1997) (near-infrared observations); Schaefer et al. (1998) (
ground-based optical observations); and Vrba et al. (1998) (UBVI observations at
the USNO).  Although no counterparts were identified in any of these observations,
the sensitivities in many cases would have been insufficient to detect the faint
galaxies which have been found in later studies by searching at the precisely known positions of
the brighter optical transients.  Except for GRB980425, where the association of
the galaxy with the GRB is still debatable, the nature of the host galaxies found so far
is uncertain.  If they are faint or star-forming
galaxies at redshifts $\sim$1, and if the four bursts in this study have similar hosts,
then it is clear from table 2 why no quiescent X-ray sources were detected; the
sensitivity would have allowed the detection of such objects only out to redshifts of
$\sim$0.2 at best.

Assuming a sensitivity of $\rm 3 \times 10^{-15} erg\, cm^{-2}\, s^{-1}$ for a deep AXAF
or XMM observation, the quiescent X-ray emission from normal, faint, or star-forming galaxies could be detected 
out to redshifts of 0.3, 0.8, and 0.2, respectively.  It is therefore possible that the
quiescent X-ray counterparts to the closer bursts could be detected.  It is also possible
that the short GRBs originate at smaller distances (as their number-intensity relation
suggests), making the X-ray detection of their
host galaxies feasible.

\section{Acknowledgments}

The work at UC Berkeley was supported by grant NAG5-1727 from the U.S. ROSAT
guest investigator program.

\clearpage

\figcaption{ROSAT HRI image of the GRB910522 field with the approximate position of the IPN error box.  \label{Fig. 1}}

\figcaption{ROSAT HRI image of the GRB920325 field with the approximate position of the IPN error box.  \label{Fig. 2}}

\figcaption{ROSAT HRI image of the GRB920406 field with the approximate position of the IPN error box.  \label{Fig. 3}}

\figcaption{ROSAT HRI image of the GRB930706 field with the approximate position of the IPN error box.  \label{Fig. 4}}

\clearpage
\begin{deluxetable}{ccccc}
\footnotesize
\tablecaption{GRB properties and ROSAT results}
\tablehead{
\colhead{ }&\colhead{GRB910522}&\colhead{GRB920325}&\colhead{GRB920406}&\colhead{GRB930706}
}
\startdata
BATSE No. & 219 & 1519 & 1541 & 2431 \nl
ROSAT Obs. No. &	US400882H & US400881H & US400879H & US400880H \nl
b$\rm^{II}$ & -2 $\arcdeg$ & -44 $\arcdeg$ & -26 $\arcdeg$ & -7 $\arcdeg$ \nl
Error box size (arcmin.$^2$) & 9.3 & 4.8 & 0.44 & 4.0 \nl 
Fluence (erg cm$^{-2}$) & $\rm 3\times 10^{-5}$ & $\rm 5\times 10^{-6}$ & $\rm 1.4\times 10^{-4}$ & $\rm 2\times10^{-5}$ \nl
Obs. time (s.) & 19,748 & 16,008 & 21,018 & 22,869 \nl
Time since GRB (yr.) & 6.50 & 5.17 & 5.50 & 4.25 \nl
Source in error box? & no & no & no & no \nl
Total number of sources in  40 $\arcmin$ FOV & 5 & 6 & 5 & 5 \nl
\it A posteriori \rm chance detection probability & 0.037 & 0.023 & 0.0018 & 0.016 \nl
N$\rm_H$, cm$^{-2}$ & $\rm 1.35 \times 10^{22}$ & $\rm 4.32 \times 10^{20}$ &
$\rm 5.34 \times 10^{20}$ & $\rm 1.92 \times 10^{21}$ \nl
3 $\sigma$ flux upper limit (brems.), erg cm$^{-2}$ s$^{-1}$ & 
$\rm 6.2 \times 10^{-13}$ & $\rm 6.4 \times 10^{-14}$ & $\rm 5.2 \times 10^{-14}$ &
$\rm 7.5 \times 10^{-14}$ \nl
3 $\sigma$ flux upper limit (blackbody), erg cm$^{-2}$ s$^{-1}$ & 
$\rm 2.8 \times10^{-13}$ & $\rm 6.8 \times 10^{-14}$ & $\rm 5.3 \times 10^{-14}$ &
$\rm 6.2 \times 10^{-14}$ \nl
3 $\sigma$ flux upper limit (power law), erg cm$^{-2}$ s$^{-1}$ & 
$\rm 3.2 \times 10^{-13}$ & $\rm 6.0 \times10^{-14}$ & $\rm 7.0 \times 10^{-14}$ &
$\rm 7.9 \times 10^{-14}$ 
\enddata
\end{deluxetable}
 
\clearpage
\begin{deluxetable}{ccccc}
\scriptsize
\tablecaption{Lower limits to GRB host galaxy redshifts and total GRB 
energies for four types of galaxies}
\tablehead{
\colhead{ }&\colhead{GRB910522}&\colhead{GRB920325}&\colhead{GRB920406}&\colhead{GRB930706}
}
\startdata
z (normal) & $>$0.0004-0.04 & $>$0.0008-0.08 & $>$0.0008-0.07 & $>$ 0.0007-0.07 \nl
E$_{\gamma}$, erg (normal) & $\rm>1\times10^{46}-1\times10^{50}$ & 
$\rm>8\times10^{45}-9\times10^{49} $ &
$\rm>2\times10^{47}-2\times10^{51} $ &
$\rm>3\times10^{46}-3\times10^{50} $ \nl

z (AGN) & $>$0.04-1.8 & $>$0.08-3.1 & $>$0.07-3.0 & $>$0.07-2.8 \nl
E$_{\gamma}$, erg (AGN) & $\rm>1\times10^{50}-3\times10^{54}$ &
$\rm>9\times10^{49}-3\times10^{54}$ &
$\rm>2\times10^{51}-8\times10^{55}$ &
$\rm>3\times10^{50}-1\times10^{55}$ \nl

z (faint) & $>$0.02-0.1 & $>$0.04-0.2 & $>$0.04-0.2 & $>$0.04-0.2 \nl
E$_{\gamma}$, erg (faint) & $>3\times10^{49}-1\times10^{51}$ &
$\rm>3\times10^{49}-1\times10^{51}$ &
$\rm>7\times10^{50}-2\times10^{52}$ &
$\rm>8\times10^{49}-3\times10^{51}$ \nl

z (star-forming) & $>$0.002-0.02 & $>$0.005-0.04 & $>$0.004-0.04 & 
$>$0.004-0.04 \nl
E$_{\gamma}$, erg (star-forming) & $\rm>3\times10^{47}-3\times10^{49}$ &
$\rm>3\times10^{47}-3\times10^{49}$ &
$\rm>6\times10^{48}-7\times10^{50}$ &
$\rm>8\times10^{47}-8\times10^{49}$ \nl 

\enddata
\end{deluxetable}

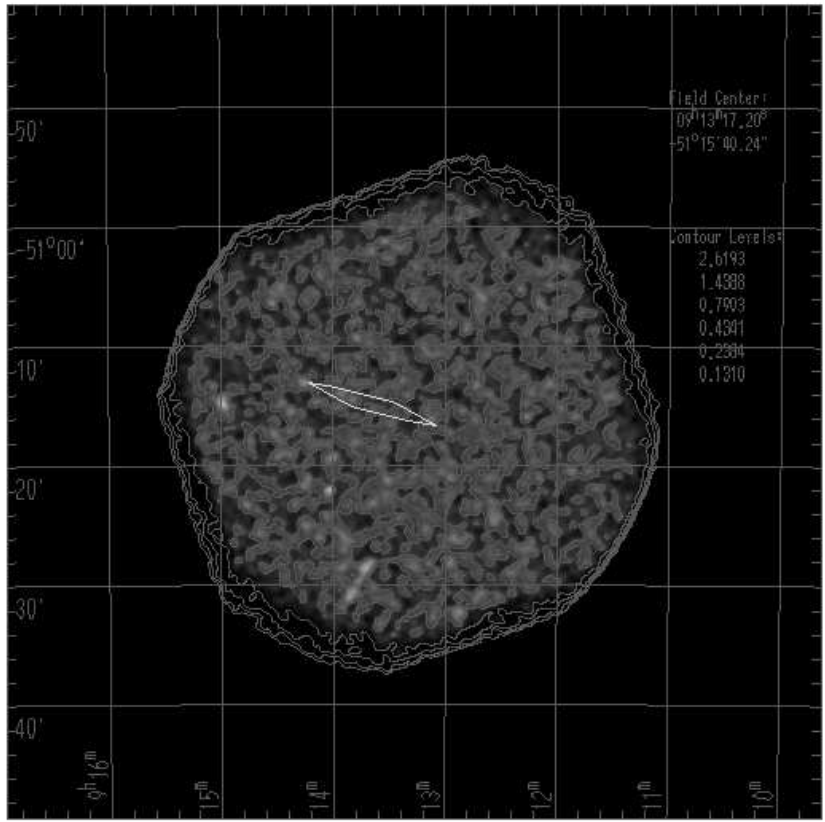
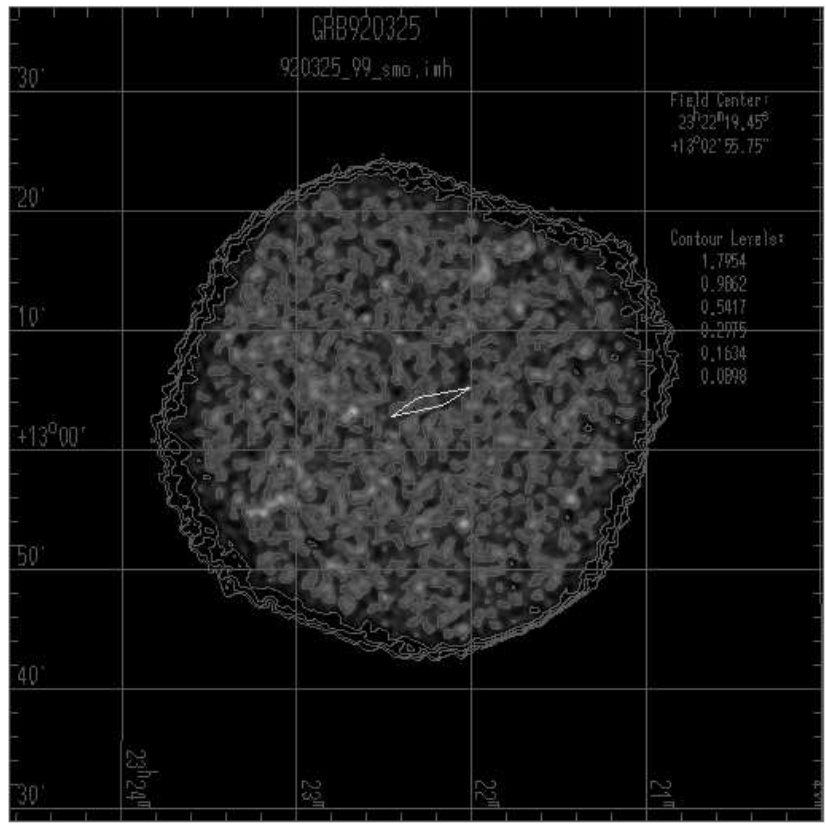
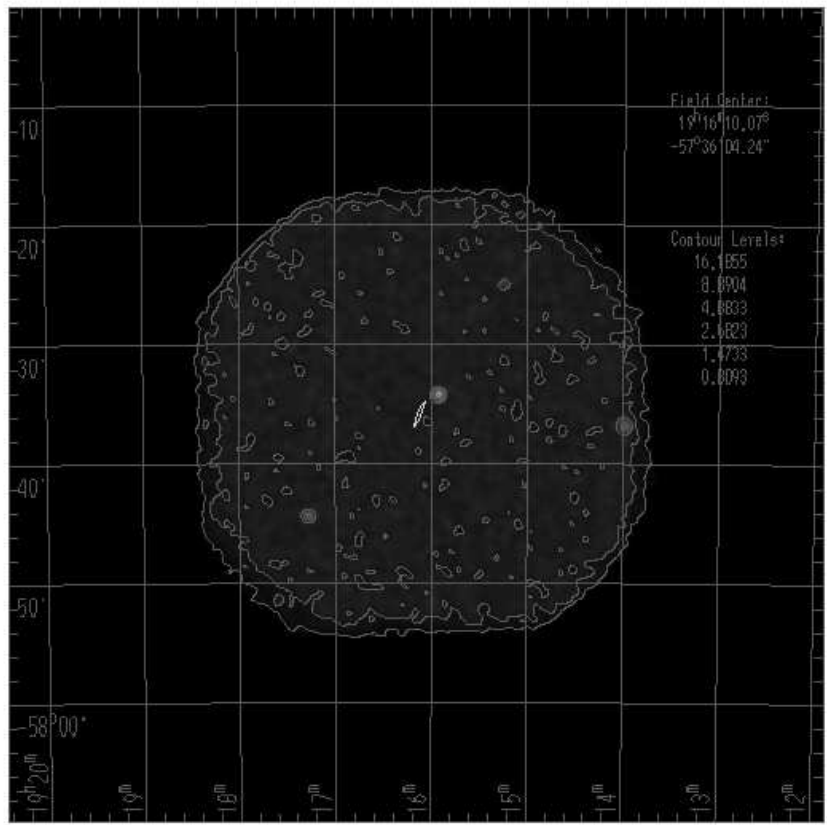
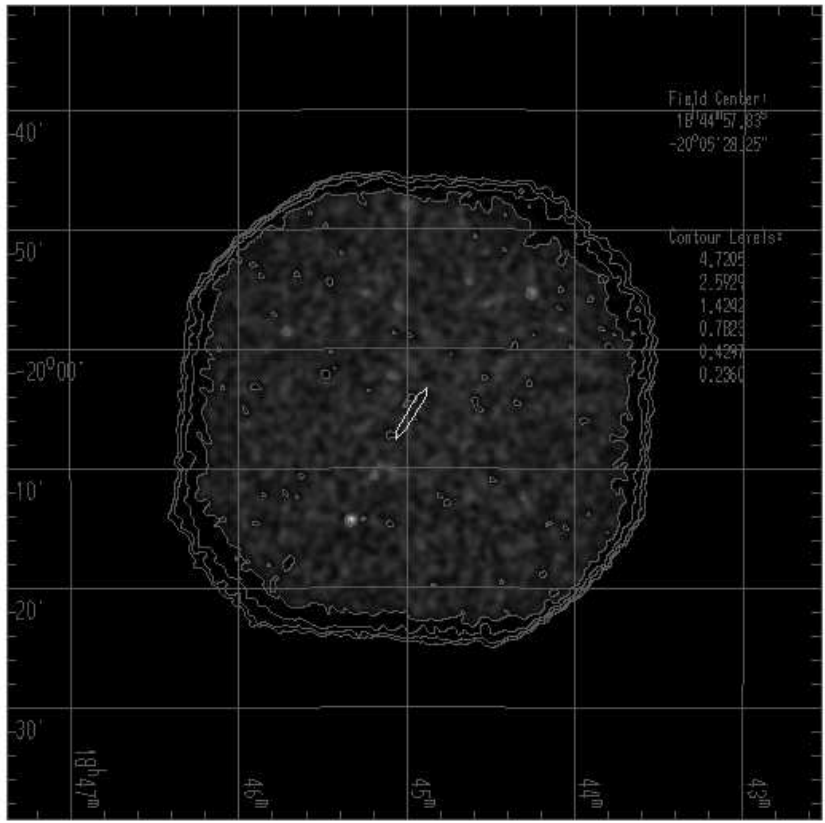
\end{document}